\documentclass[aps,twocolumn,shopacs,floats,superscriptaddress, nofootinbib,prx]{revtex4}
\usepackage{graphicx}
\usepackage{epsf}
\usepackage{epsfig}
\begin{document} 
\preprint{}
\preprint{KEK-TH 1088}
\title{Novel reconstruction technique for New Physics processes 
with initial state radiation}
\author{Johan Alwall}
\affiliation{SLAC, MS 81, 2575 Sand Hill Road, Menlo Park CA 94025, USA  }
\author{K. Hiramastsu}
\affiliation{IPNS, KEK and The Graduate University of Advanced Studies[SOKENDAI], Oho 1-1  305-0801, Japan}
\author{Mihoko M. Nojiri}
\affiliation{IPNS, KEK and The Graduate University of Advanced Studies[SOKENDAI], Oho 1-1  305-0801, Japan}
\affiliation{IPMU, 5-1-5 Kashiwanoha, Kashiwa, 277-8568, Japan }
\author{Yasuhiro Shimizu }
\affiliation{IIAIR, Tohoku University, Aobaku, Sendai 980-8578, Japan}
\date{\today}
\begin{abstract}
At hadron colliders, the production of heavy new particles is
associated with additional quarks and gluons with significant
transeverse momentum. The additional jets complicates the
reconstruction of new particle masses. Taking gluino pair production
and decay at the Large Hadron Collider as an example, we
develop a novel technique to reduce these effects, and to reconstruct a
clear kinematical endpoint for the gluino decay products.
\end{abstract}
\maketitle 
Although the Standard Model very well describes the interactions of
elementary particles, the Higgs boson, the particle responsible for
the electroweak symmetry breaking, is associated with a hierarchy
problem. Many models have been proposed to solve this problem, some of
which can also account for the dark matter inferred by astronomical
observations. Examples of such models are Supersymmetric models with R
parity and Little Higgs models with T parity.  These models predict
the existence of new colored particles around 1 TeV, as well as a
stable lightest new particle.


In supersymmetric models, squarks and gluinos will be 
copiously produced in the ATLAS and CMS experiments at the LHC \cite{ATLASCMS}. 
By looking at kinematical 
distributions of the decay products, the masses of squarks and gluinos 
can  be reconstructed. However, since the production and 
decay processes involve particles charged under QCD,  
initial and final state QCD radiation will complicate this
reconstruction \cite{NLO}.  
 
The problem of initial state radiation (ISR) is generic for all new
physics signatures involving multiple hard jets in the final state. It
reduces the precision on mass determinations in hadronic
channels. Understanding of jet emission associated with hard
processes is becoming one of the central issues of the LHC
phenomenology.
 
In Supersymmetric models, the effect of initial state radiation is
especially severe for the pair production of gluinos. The lowest order
process $pp\rightarrow \tilde{g}\tilde{g}$ is dominated by $g g
\rightarrow \tilde{g}\tilde{g}$. When both gluinos decay
into $q\bar{q}\tilde{\chi}^0_1$, it is possible to determine
$m_{\tilde{g}}$ and $m_{\tilde\chi^0_1}$ from the jet
distributions. In particular, it was recently realized that the gluino
and LSP masses can be simultaneously reconstructed by looking at the end
point of $M_{T2}$ distributions \cite{Barr:2003rg,Cho:2007qv}, if we can correctly identify two jet
pairs that arise from gluino decays.  However, by picking up jets
coming from initial state radiation, the end point is significantly
smeared, leading to increased uncertainty in the end point
determination.

The aim of this paper is to propose a new jet selection method to
reduce the misreconstruction due to the initial state radiation.
Systematic studies of the effect of additional QCD radiation has recently
become possible thanks to Monte Carlo (MC) simulations which include
the effect of hard parton emission from inital state quarks and gluons
in new physics processes \cite{Alwall:2008qv}.
 
Processes involving initial state radiation can be simulated using
parton shower (PS) MCs such as PYTHIA or
 HERWIG. They generate soft
and collinear parton shower emissions in association with hard two-to-two
processes. Hard parton emissions, on the other hand, are not correctly
described by this approximation. There, matix element (ME) calculations
are required to correctly predict distributions. Various matching
techniques have been developed to remove double counting between PS
emission and ME contributions and account for Sudakov suppression
effects \cite{Matching}. These techniques are important to compare the differential
distributions of the hard jets with theoretical predictions.  Detailed
comparisons of various event generators with PS-ME matching have been
performed \cite{Alwall:2007fs}, showing that the matching schemes used
in Sherpa, MadGraph, ALPGEN, and ARIADNE are consistent each other,
although the algorithms are slightly different.

For SUSY processes, PS and ME matching is available in
MadGraph/MadEvent \cite{Alwall:2007st} matched with Pythia
\cite{Sjostrand:2006za}. We here use it to study the effect of initial
state radiation in sparticle mass reconstruction at the LHC.  We have
generated sparticle production events with up to one additional
quark/gluon.  The matching scheme used is called \lq\lq$k_T$-jet MLM"
scheme, described in detail in \cite{Alwall:2008qv}. In
this scheme, the final state partons in an event are clusted according
to the $k_T$ jet algorithm to find the equivalent parton shower
history of the event. The smallest $k_T$ value is restricted to be
above a cut off scale $Q_{\rm cut}^{\rm ME}$. After showering, the
final state partons are clustered into jets using the $k_T$ jet
algorithm with a cutoff scale $Q_{\rm match} > Q_{\rm cut}^{\rm
ME}$. The event is rejected unless each jet is matched to a parton,
except for the highest multiplicity sample.  
Events from the $pp\rightarrow \tilde{X} \tilde{X}'$ contribution to
the matched sample, where $\tilde{X}$ and $\tilde{X}'$ is either a
gluino or a squark, are here referred to as \lq exclusive' events,
while events from the highest multi-jet sample (in our case
$\tilde{X}\tilde{X}j$), which contains hard emissions as well as
resolvable PS effects, are called \lq inclusive'.  In this paper we
take $Q_{\rm match}$=60~GeV and $Q_{\rm cut}^{\rm ME}=40$~GeV, so \lq
inclusive' events have additional hard radiation with $p_T > $60~GeV.
  

For demonstration purposes, we focus in this paper on the study of
gluino pair production. For simplicity we force the gluinos to decay
into $u\bar{u}\tilde{\chi}^0_1$. In this context, the exclusive sample
is events with four partons + 2 LSP at parton level, and the inclusive
sample is events with five partons + 2 LSP. Squark masses are taken to
be high enough so that their production can be ignored. We take an
MSSM point with $m_{\tilde{g}}=685$~GeV and $m_{\tilde q}=1426$~GeV,
and $m_{\tilde{\chi}^0_1}= 101.7$~GeV.  We generate $1.21\times 10^5$
events. After removing squark gluino associated production (where an on
shell squark decays into $\tilde{g}q$ to generate the $\tilde{g}\tilde{g}q$
final state), $1.07 \times 10^5$ events remains. The cross section is
2.5 pb, so the number of generated events corresponds to $40~{\rm
fb}^{-1}$.
 
The generated events are then simulated by the toy detector 
simulator AcerDET \cite{RichterWas:2002ch} with the jet reconstruction 
tool Fastjet \cite{Cacciari:2006sm}.  
In AcerDET, the phase space is divided into cells with $(\Delta\eta,
\Delta \phi) = (0.1, 0.1)$, and the momenta of hadrons, electrons, and
photons passing within one cell are 
summed to imitate the  energy deposit in a calorimeter cell. 
The cell energy deposits are interfaced to Fastjet for jet reconstruction
as  massless particle momenta. 
 The momenta of the reconstructed jets are then 
 smeared as $p_{\rm smear} = (1+\delta) p_{\rm jet}$, 
 with the energy resolution $\delta= 0.5(1)/\sqrt{E_{\rm jet}}$ 
  assumed in the barrel(forward) direction with $\vert \eta\vert<(>)3.2$.
 We use the Cambridge-Achen algorithm with $R=0.4$ 
 in  this paper.
\footnote{In principle, the jet resolution should depend on 
the reconstruction algorithm and $R$.  In addition, energy smearing
and momentum smearing may not be the same. The smearing we introduced
in this work is only for illustrative purposes.}
The missing $p_T$ was calculated from the reconstructed objects after
smearing. 

\begin{figure}
\begin{center}
\epsfxsize=0.23\textwidth\epsfbox{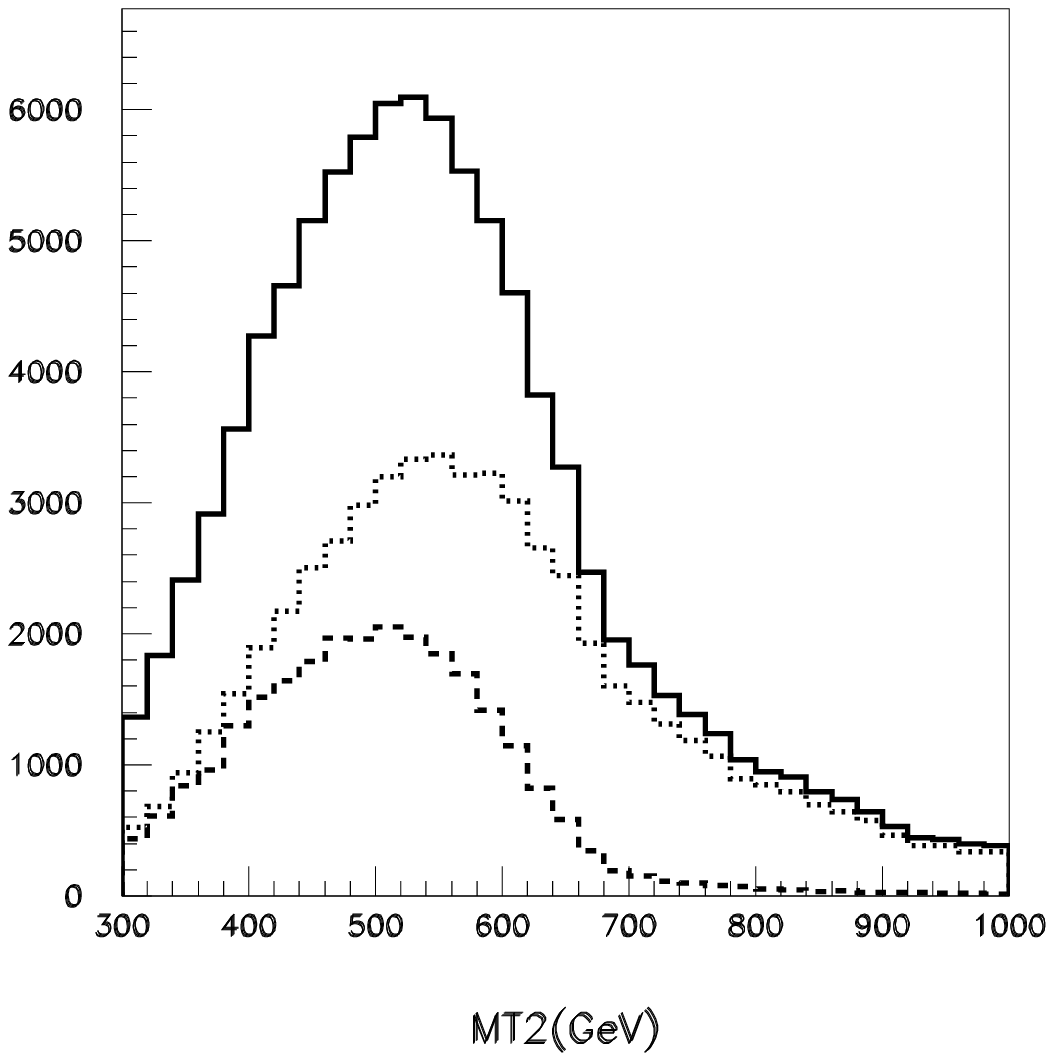}
\epsfxsize=0.23\textwidth\epsfbox{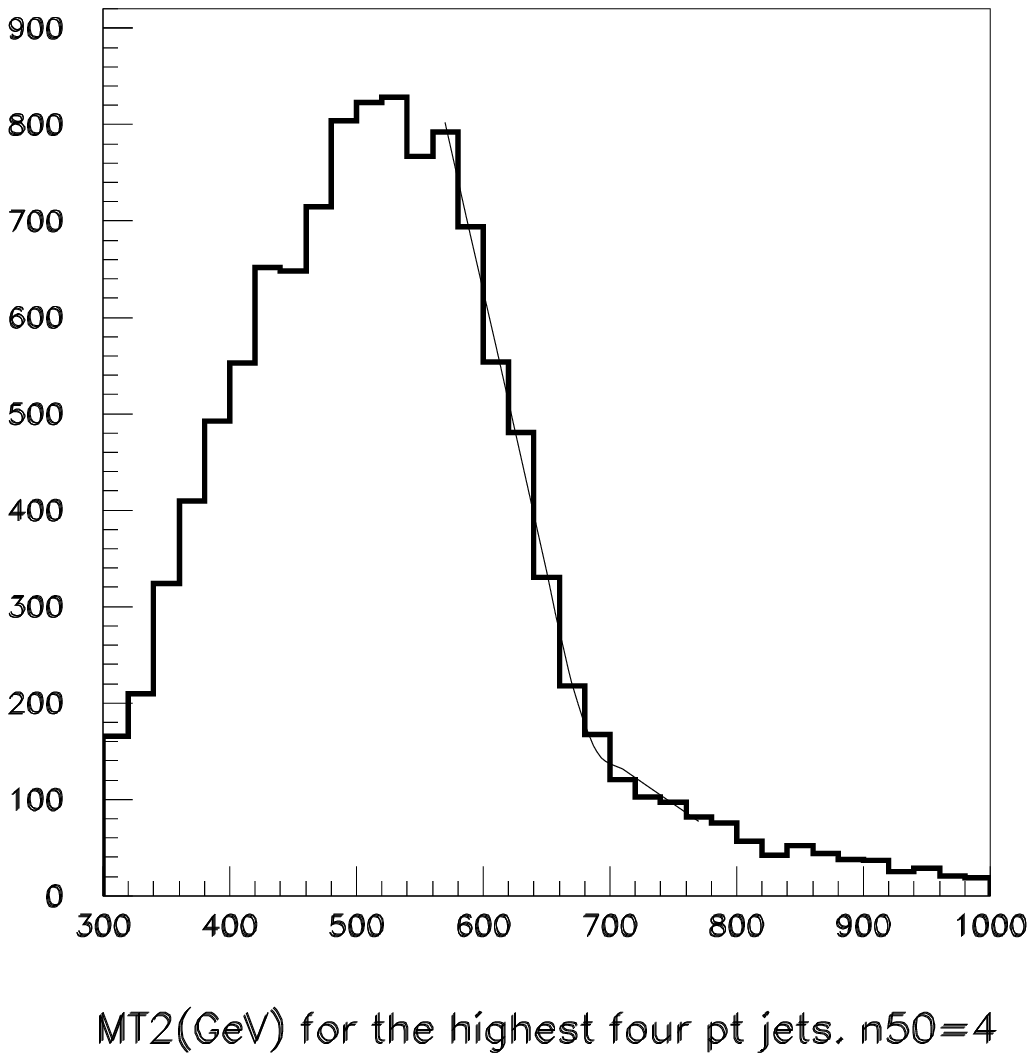}
\end{center}
\caption{a) (left)  The $m_{T2}$ distribution calculated 
from the four highest $p_T$ jets, using all events. 
b) (right) The $M_{T2}$ distribution
of events with exactly 4 jets with $p_T>50$~GeV.
}\label{fig:mt2jetdif}
\end{figure}

In Fig.~\ref{fig:mt2jetdif} a), we show the $M_{T2}$ distribution
calculated using the four 
hardest jets. The $M_{T2}$ observable is 
calculated from two visible momenta 
$p^{\rm vis}_{1}$, $p^{\rm vis}_{2}$, 
a test LSP mass $m^{\rm test}_{\chi}$, and  test LSP momenta
$p_{1\chi}$  and $p_{2\chi}$ satisfying the constraint 
$p_{1\chi}+ p_{2\chi}=p_T^{miss}$ 
as follows,
\begin{eqnarray}
&&M_{T2}= \min_{p^T_{1\chi}+p^T_{2\chi}= p_{\rm miss}^{T}} 
\cr 
&&
\left[\max \left(M_T(p_{1}^{\rm  vis},p^T_{1\chi}, m^{\rm test}_{\chi}), 
 M_T(p_{2}^{\rm  vis},p^T_{2\chi}, m^{\rm test}_{\chi}\right)\right]. 
 \cr
&&
\end{eqnarray}

If both $p^{\rm vis}_{1}$ and $p^{\rm vis}_{2}$ are the momenta of the
sum of a visible gluino decay products and $m^{\rm test}_{\chi}$ is taken as
the LSP mass, the $M_{T2}$ endpoint should coincide with
$m_{\tilde{g}}$.  It has also recently been pointed out that the
$M_{T2}$ end point, taken as a function of the test mass $m^{\rm
test}_{\chi}$, shows a kink at the true LSP mass.  Therefore, the LSP mass
and gluino mass can be determined simultaniously. These ideas have got
significant attention in the literature, and various extensions are
beeing studied.

Experimentally, we cannot know from which parent particle a jet
arises.  In Fig.~\ref{fig:mt2jetdif} a), we have defined $p_{\rm vis}$
as follows:
\begin{enumerate}
\item We first take the two highest 
 $p_T$ jet  momenta $p_1$ and $p_2$ as seeds. 
\item  We then calculate $M_{T2}$ for 
 the combinations 1)  $(p_1^{\rm vis}, p_2^{\rm vis}) 
=( p_1 +p_3, \ p_2+p_4 ) $  and 2) $(p_1+p_4, \ p_2+p_3)$
  giving $M^{1(2)}_{T2}$ and take the minimum, 
  $M_{T2} =\min(M^{1}_{T2}, M^2_{T2})$.  
\end{enumerate}
 
In the Fig.~\ref{fig:mt2jetdif} a) we show the $M_{T2}$
distribution of the four highest $p_T$ jets paired as above for the
events with $n_{50}= ($number of jets with $p_T>50$~GeV) $\ge 4$ as the solid
line.  The distribution does 
not show a clear end point. This is due to the effects of the initial
state radiation. To show this, we plot the distributions of the
exclusive and inclusive samples separately in the same figure. A
dashed line shows the distribution of exclusive events, i.e., events
in the matched sample which have no resolvable QCD radiation above 60
GeV (4-parton events). The dotted line shows the distribution for the
inclusive sample, i.e., the contribution from five parton events. The
exclusive sample has an end point at the correct gluino mass, since it
does not contain hard additional jet activity besides that coming from
gluino decays. The ratio $N({\rm 5 \ parton})/N({\rm 4 \
parton})$ is here 1.4.  The fraction of the events with
additional partons after the matching is larger for gluino pair
production compared with squark pair production. For example, at the
SPS 1a point, where the gluino mass is 595~GeV and squark mass is around
530~GeV, The ratio $N(\tilde{X}\tilde{X}j)/N(\tilde{X}\tilde{X})$ is
1.47 for $X=\tilde{g}$ and 0.81 for $X=\tilde{Q}$ ($=
\tilde{u},\tilde{d}, \tilde{c}, \tilde{s}$ and its charge conjugates)
for a 60~GeV jet resolution scale.

\begin{figure}
\epsfsize=0.2\textwidth\epsfbox{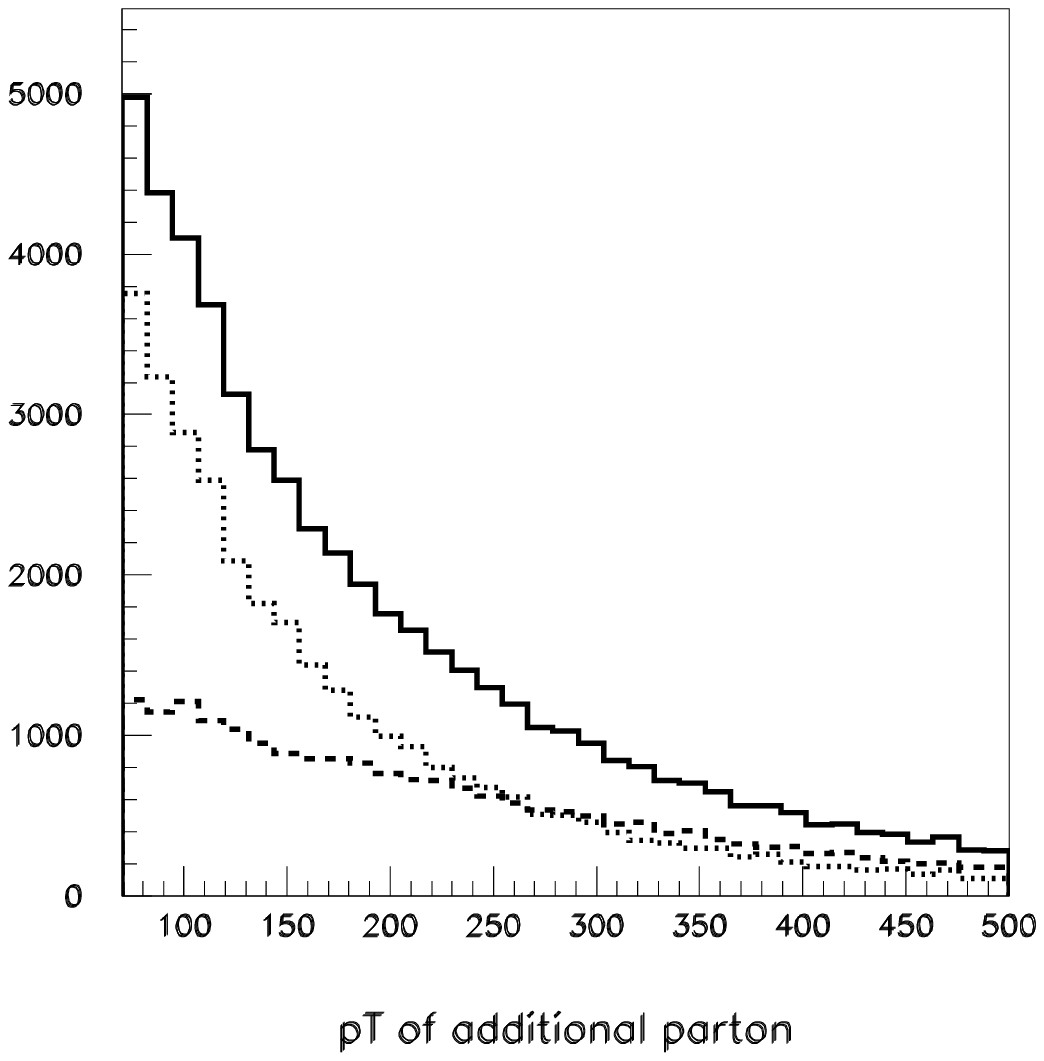}
\epsfxsize=0.2\textwidth\epsfbox{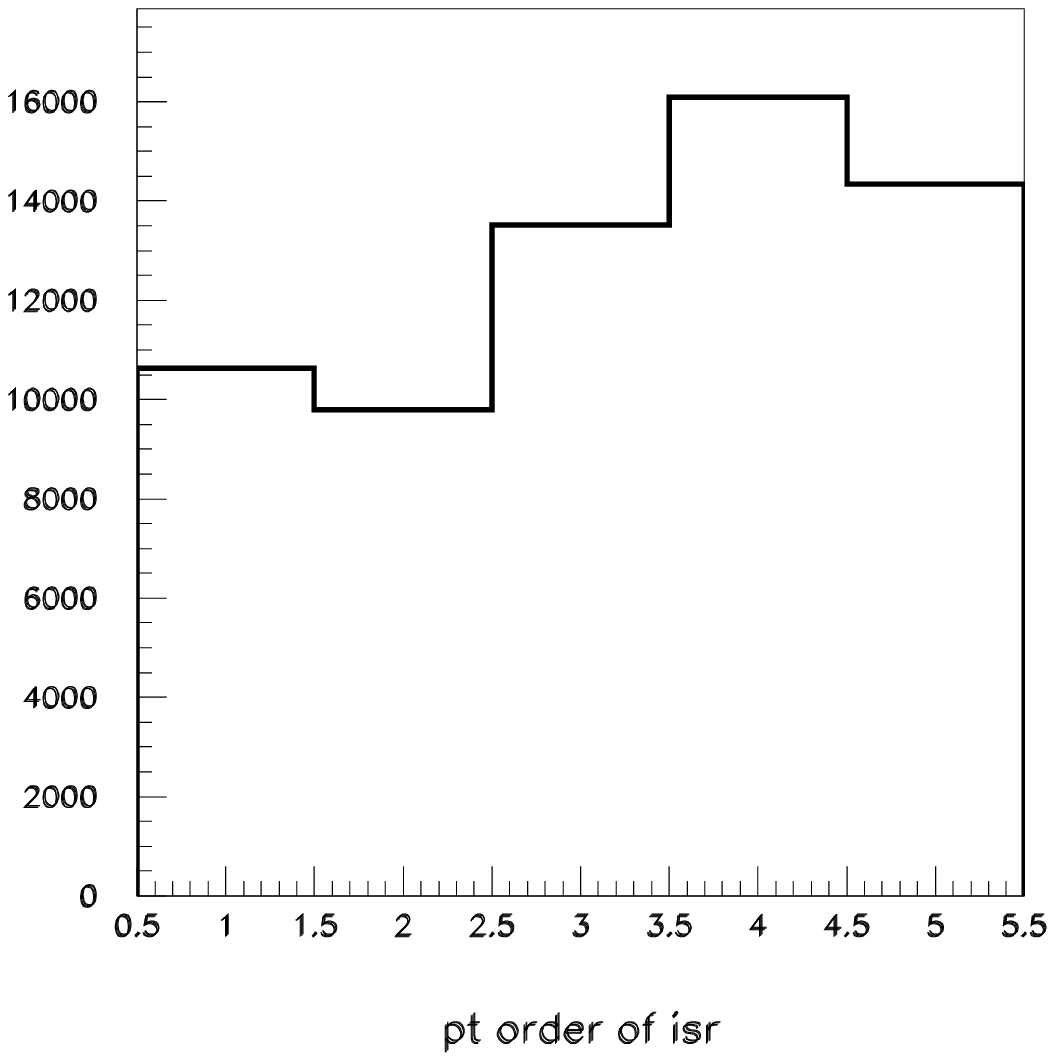}

\caption{a)(left $p_T$ distribution of the additinal parton for $pp\rightarrow \tilde{g}\tilde{g} j$. The dotted(dashed)  line shows the distribution for $j=g(q)$. 
b) (right)  $p_T$ order of the ISR parton among the five parton of the 
inclusive sample.
  } 
\label{fig:isrpartona}
\end{figure}

There are several reasons why the smearing of the end point due to
initial state radiation is here relatively large. In the simulation,
the gluino is forced to decay into a three body final state.  The
typical $p_T$ of the partons from gluino decay is therefore $\sim
m_{\tilde{g}}/3$. On the other hand, the $p_T$ of the initial state
radiation is not small, in average O(100)~GeV, because the produced
gluino is heavy.  The $p_T$ distribution is shown in
Fig.~\ref{fig:isrpartona} a).
The $p_T$ of the additional parton is large in average, and often
larger than one of the partons from gluino decay, as can be seen in
Fig.~\ref{fig:isrpartona} b) where the position of of the initial
state parton among all partons, ordered in $p_T$, is plotted.  The
probablity that ISR parton is the 5th, softest, parton is only about
22\% of all five parton events. 

One may recover the clean end point by requiring exactly 4 jets with $p_T>50$~GeV in the final 
state (Fig.~\ref{fig:mt2jetdif} b).
However, this selection is not practical for general MSSM model
points. The reason is that we expect the decay branching ratio 
into heavier neutralino or chargino, 
$\tilde{g}\rightarrow \tilde{\chi}^0_i q\bar{q}, \tilde{\chi}^+ q\bar{q}'$, 
to be large,  where $\tilde{\chi}_i^0, \tilde{\chi}^+_i$ further decays 
into jets and leptons. The branching ratio for both of the 
gluinos to  decay into 2 jets and LSP may be small, in which case
this cut would reduce the statistics significantly.

A better solution is  obtained by taking into account the existence of 
additional ISR jets in the analysis. 
 Given the high probability to have ISR jets,  we 
should regard  the process we are interested in 
 as a  {\bf \it five-jet system} rather than the
four-jet system expected at the lowest order.  We therefore propose 
the  reconstruction of a five jet distribution, rather than four jet
distribution hitherto considered.

For this purpose, we define $M_{T2}(i)$ ($i=1..., 5$) , where $M_{T2}(i)$ is calculated 
from the five highest $p_T$  jets, excluding the $i-$th highest $p_T$ jet. 
\begin{equation}
M_{T2}(i) = M_{T2} (p_1, ... , p_{i-1}, p_{i+1}, ...p_5) 
\end{equation}

\begin{figure}
\epsfxsize=0.2\textwidth\epsfbox{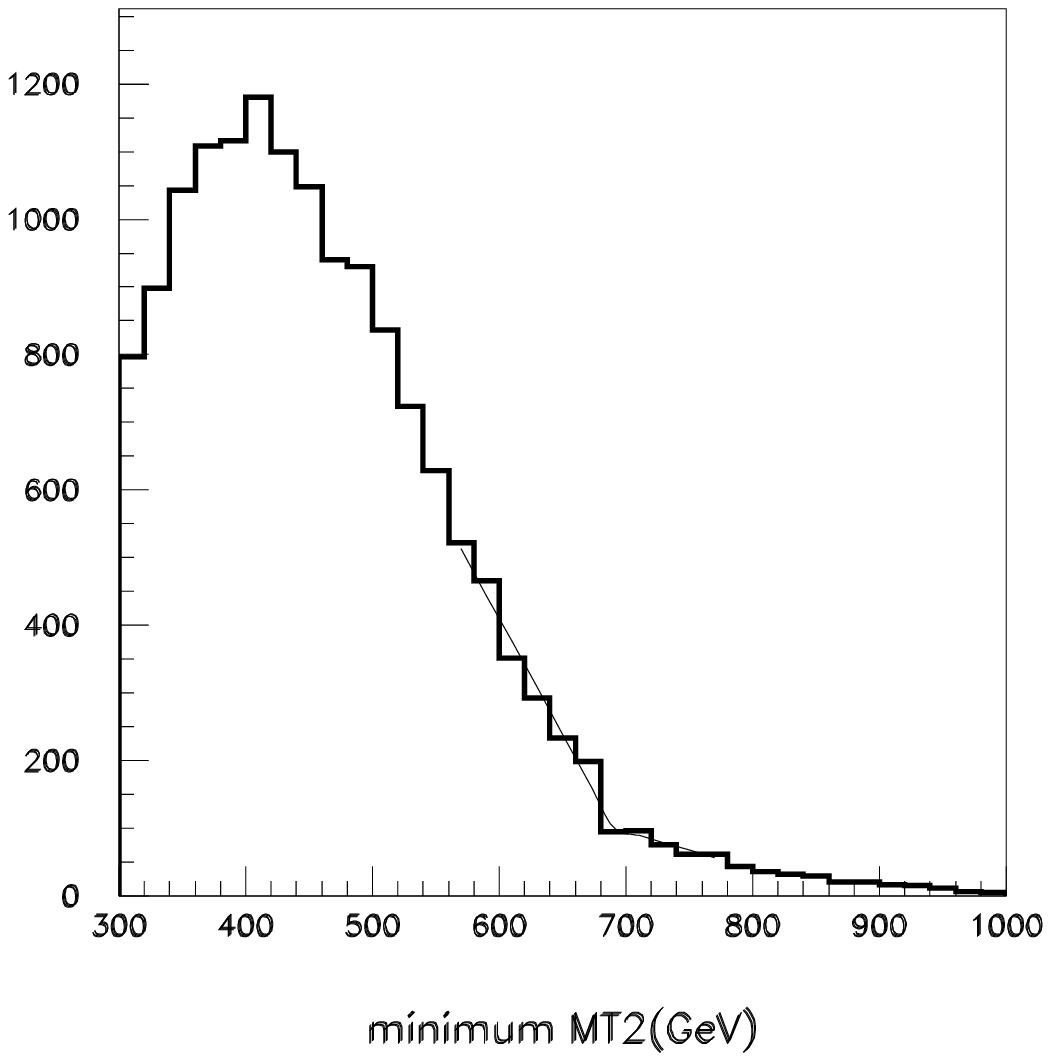}
\epsfxsize=0.2\textwidth\epsfbox{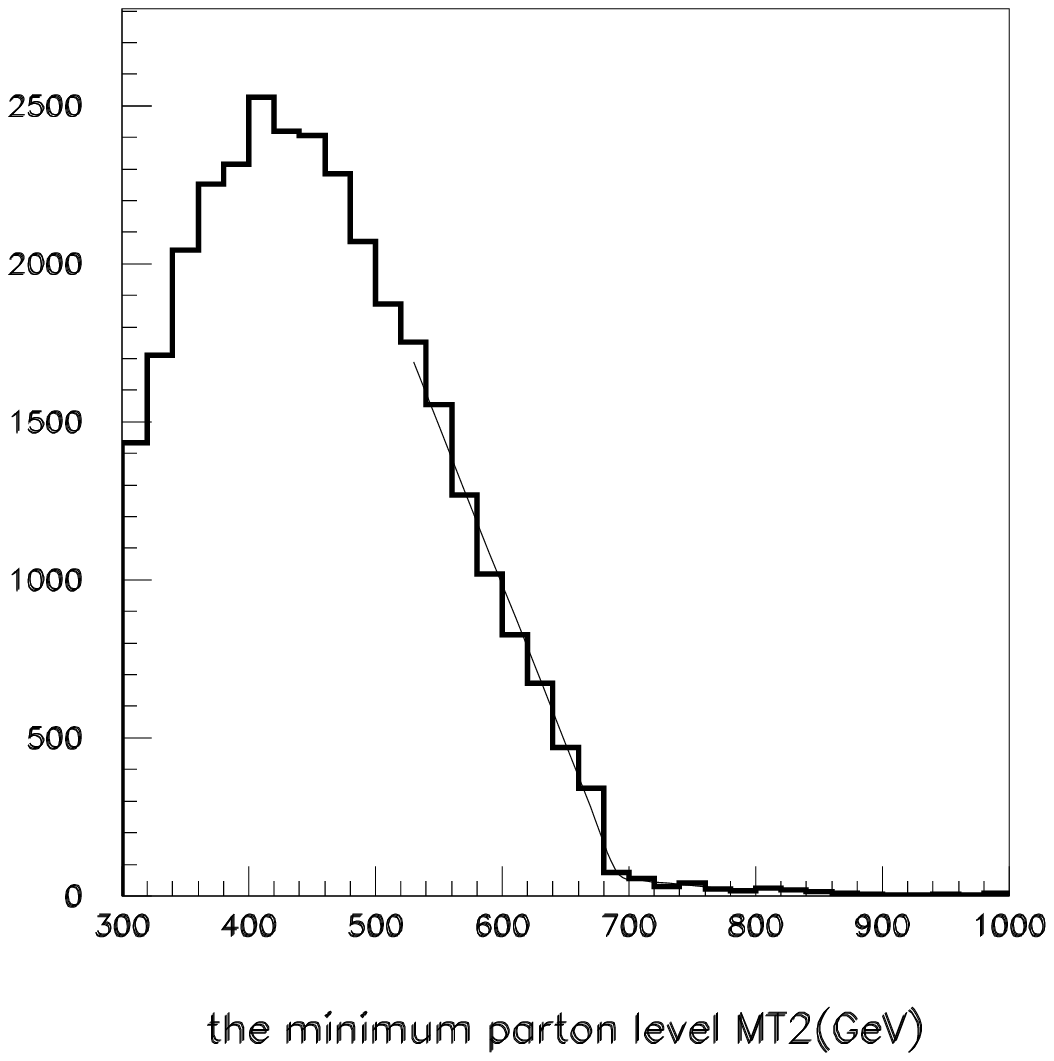}
\caption{a) (left) Distribution of $M^{\rm min}_{T2}$ at jet level for the events 
with $n_{50}\ge 5$ and $i_{\rm min}\ge 3$. b) (right) parton level $M^{\rm min}_{T2}$ distribution 
for the  5 parton sample.}
\label{fig:mt2min}
\end{figure}

We find that the problem arising from the ISR is  significantly 
reduced if we look at the
$M^{\rm min}_{T2}$ distribution, where 
\begin{equation}
M^{\rm min }_{T2}\equiv \min_{i=1,..5} (M_{T2}(i)).
\end{equation} 
At the parton level,  
$M^{\rm min}_{T2}$ contains at least one correct parton combination, and
therefore 
$M^{\rm min}_{T2}<M_{T2}^{\rm end}.
$
The  jet level distribution is shown in Fig.~\ref{fig:mt2min} a)
for events with $n_{50}\ge 5$ and 
$i_{\rm min} \ge 3$, where $i_{\rm min}$ satifsy $M_{T2}(i_{\rm min})=M^{\rm min }_{T2}$. 
Note that the events in 
Fig.~\ref{fig:mt2min} a) are statistically independent from those in
Fig.~\ref{fig:mt2jetdif} b).  

In the figure, we do not include the events  with  $i_{\rm min}=1, 2$,  because $M_{T2}(1)$, and $M_{T2}(2)$
tend to be softer  than the others since the removed jet has a high $p_T$. 
The distribution of $i_{\rm min}=1,2$ appears to be smeared
and the end point is not well determined.
The parton level  $M^{\rm min}_{T2}$ distribution  for the  five parton sample 
is  given in Fig.~\ref{fig:mt2min}~b).
 
\begin{figure}
\vskip 0.5cm
\includegraphics[width=3cm, angle=90]{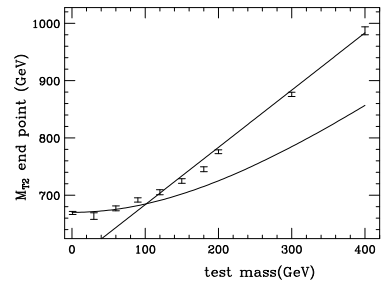}
\caption{The test mass dependence of the $M^{\rm min}_{T2}$ end point.  }
\label{fig:testmass}
\end{figure}

We fit a $f(x)$ to the $M^{\rm min}_{T2}$ distributions, where 
$f(x)=\Theta(x-M^{\rm end}) [a_1(x-M^{\rm end})+b] $$+\Theta(M^{\rm end}-x) [a_2 (x-M^{\rm end})+b] 
$
to see if the end points are recovered correctly.  The fitted end point 
 $M^{\rm end}_{T2}$ is 692.3 $\pm$  1.2~GeV at  
parton level for 5 parton events with  $i^{\rm parton}_{\rm min}\ge 3$. 
This should be compared with $M^{\rm end}_{T2}$ for 
the parton level $M_{T2}$ distribution calculated using the 
partons coming from the $\tilde{g}$ decay,  $694.1\pm 0.5$~GeV. The 
$M^{\rm end }_{T2}$ end point at jet level is  691.5$\pm$ 3.9~GeV for events with 
$n_{50}\ge 5$ and $i_{\rm min}\ge 3$. The $M^{\rm end}_{T2}$ calculated for 
$n_{50}=4$ events is  692.3 $\pm$ 2.4~GeV. 
These values  are consistent with the input gluino mass
$m_{\tilde{g}}=685$~GeV.  
The central value  depends slightly on the fit region, and careful
study is needed to identify systematical errors.  
As a cross check, we also study the distributions with 
$k_T$ and anti-$k_T$  jet reconstruction algorithms\cite{Cacciari:2008gp} and find the 
results are consistent with these values. 
We have also checked the test mass dependence on the end point. 
In Fig.~\ref{fig:testmass}, the bars  show the fitted 
$M^{\rm min }_{T2}$ end points and the error for  the events $n_{50}\ge 5$ 
and $i_{\rm min}\ge 3$. 
The end points are close to the
expected end point values shown in the solid lines.%


Several comments are in order.  First, to study 
longer cascade decay chains, we have to consider 
processes with more than four final state jets.  For 
example,  when $\tilde{g}$ decays into 
heavier inos, there may be additional jets coming 
from the neutralino and chargino decays. With 
GUT relations for the gaugino masses,
jets coming from the end of the cascade decay chains are softer than those 
coming directly from the gluino decay.  Our assumption 
that one of the five highest $p_T$ jets 
arises  from ISR is therefore reasonable. 
In such a case, one may use 
the inclusive definition of $M_{T2}$ proposed in \cite{Nojiri:2008hy,Nojiri:2008vq}.
In this approach, $p_{\rm vis}$ is defined using all jets and 
leptons in the final state so that they satisfy 
{\small  $p^{(1)}_{\rm vis}=\sum_i  p^{(1)}_i$, $ p^{(2)}_{\rm vis}=\sum_i  p^{(2)}_i$}
where {\small $p^{(1)}_i$} and {\small $p^{(2)}_i$} are  jets or  lepton momenta 
which satisfy
{\small $d(p^{\rm (1)}_{\rm vis}, p^{\rm (1)}_i) <d(p^{\rm (2)}_{\rm vis}, p^{\rm (1)}_i)$, 
$d(p^{\rm (2)}_{\rm vis}, p^{(2)}_i) <d(p^{\rm (1)}_{\rm vis}, p^{\rm (2)}_i)$},
with $d$ being some distance measure.  The inclusive $M_{T2}$ can also
be used to determine squark and gluino masses when $m_{\tilde{q}}>
m_{\tilde{g}}$. 
%
In the inclusive definition of $M_{T2}$, we may 
again remove one of the leading five jets, use 
hemisphere reconstruction to define $p_{vis}$, and
then calculate $M^{\rm min}_{T2}$.  This method should be 
useful to reduce the contamination from squark-gluino. 
 
\begin{figure}{
\epsfxsize=0.23\textwidth\epsfbox{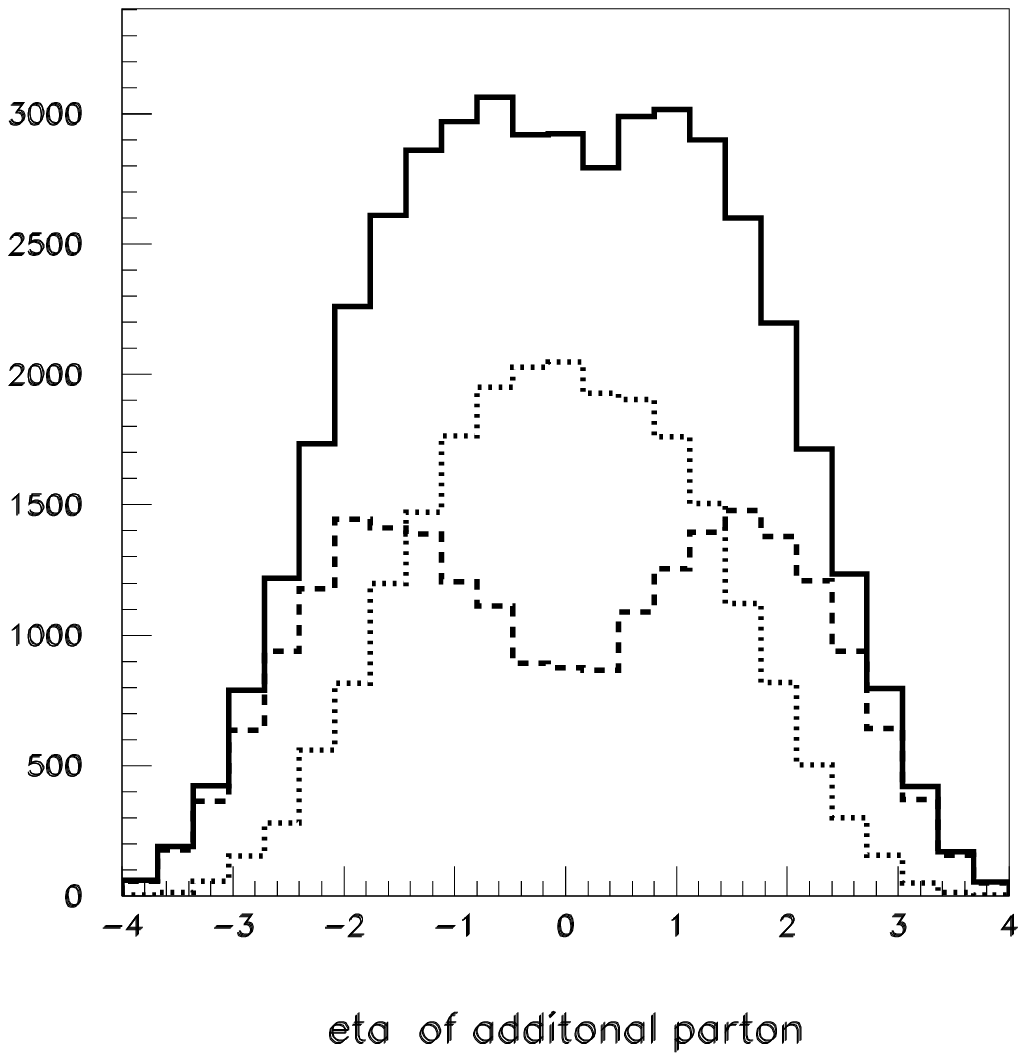} 
\epsfxsize=0.23\textwidth\epsfbox{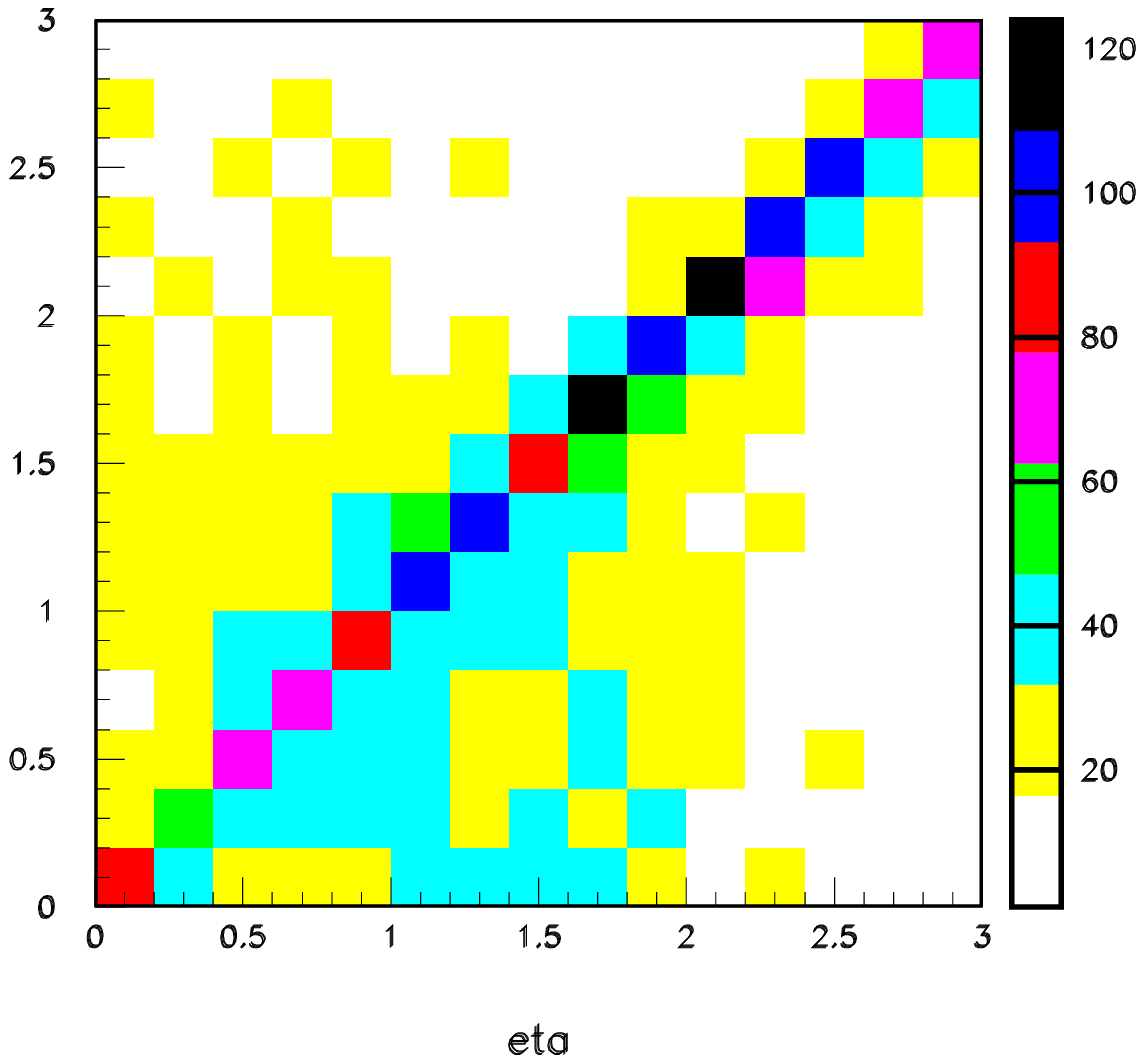}
\caption{(a) (left) $\eta$ distribution of  ISR  quark and gluon 
of $\tilde{g}\tilde{g}j$ production (solid line) for those with 
$p_T> 100$~GeV. The dashed line is for  
ISR quark distribution and dotted line is for ISR gluon. b)  (right) 
Correlation between $\vert \eta \vert $ of ISR parton 
and $\vert \eta\vert$ for the jets that gives $M^{\rm min}_{T2}$.  }\label{fig:isrparton}}
\end{figure}

Second, contrary to naive intuition, the additonal ISR jet cannot
be removed by excluding jets with high $\eta$ or low $p_T$ from the
kinematical reconstructions.  We have seen already that the average
$p_T$ of the that additional parton is rather high.  In addition, the ISR
jets are central.  In Fig.~\ref{fig:isrparton} a) , we show the $\eta$
distribution of the additinal parton for $\tilde{g}\tilde{g}j $. We
see that gluino ISR is almost central, while quark ISR is rather forward. However, 
they tend to be at high energy, as can be seen in Fig.~\ref{fig:isrpartona} a). 
For further details we refer to \cite{inprogress}.

Finally, the proposed method makes it possible to select ISR jets, by
requiring additional cuts to $M^{\rm min}_{T2}$. For an event near the
$M_{T2}$ end point, the removed jet has a higher probability
to be the ISR jet. The probability that a different jet
combination is correct is small, because the correct value has to be in
the narrow range $M^{\rm min}_{T2}< M^{\rm true}_{T2}$ $<M^{\rm
end}_{T2}$. To check this, we study the nature of the removed parton
that gives $M^{\rm min}_{T2}$ in parton level.
Among the 5 parton events generated, only 
29\% of the partons that give the  
 $M^{\rm min}_{T2}$ is the ISR parton, 
 if no restriction is applied to $M^{\rm min}_{T2}$.
This fraction increases to 44\% for
events with  $M^{\rm min}_{T2}> 500$~GeV, 29\% of total events. 
In Fig.~\ref{fig:isrparton} b), we show a 2-dimentional plot 
where the $x$-axis is the $\vert \eta\vert$ of the ISR parton  
and $y$-axis is $\vert\eta\vert$ of the jet that gives 
$M^{\rm min}_{T2}$.  The correlation is especially good 
for $\vert\eta\vert> 2$, roughly 65\% for the forward jets that match correctly to the
 ISR parton within $\vert\Delta\eta\vert<1$. This is  because the jets from gluino decay mostly 
goes to the central regions. This shows it is possible to study 
forward ISR jet distributions associated with the hard process. 

In this paper we have seen that ISR is an important feature in
$\tilde{g}$ production at the LHC, and we have developed a method to
reduce the effect of ISR production on the gluino mass
determination. This method can also be used to
identify initial state radiation jets. The method can be 
applied for any new physics processes. 
The application of this method to other SUSY
processes as well as other models for new physics, and to the
corresponding Standard Model backgrounds, 
will be discussed in following publications \cite{inprogress}.

\end{document}